\documentclass[aps,pra,twocolumn]{revtex4}

\usepackage{graphicx,latexsym,color}

\def\be{\begin{equation}}

\def\ee{\end{equation}}

\def\bea{\begin{eqnarray}}

\def\eea{\end{eqnarray}}

\def\bi{\begin{itemize}}

\def\ei{\end{itemize}}

\begin{document}
\hbadness = 10000

\title{UV light-induced atom desorption for large rubidium and potassium 

magneto-optical traps}

\author{C. Klempt, T. van Zoest, T. Henninger, O. Topic, E. Rasel, W. Ertmer, and J. Arlt}
\affiliation{Institut f\"ur Quantenoptik, Universit\"at Hannover,
Welfengarten 1, D-30167 Hannover, Germany}

\date{\today}

\begin{abstract}
We show that light-induced atom desorption (LIAD) can be used as a flexible atomic source for large $^{87}Rb$ and $^{40}K$ magneto-optical traps. The use of LIAD at short wavelengths allows for fast switching of the desired vapor pressure and permits experiments with long trapping and coherence times. The wavelength dependence of the LIAD effect for both species was explored in a range from $630\ nm$ to $253\ nm$ in an uncoated quartz cell and a stainless steel chamber. Only a few $mW/cm^2$ of near-UV light produce partial pressures that are high enough to saturate a magneto-optical trap at $3.5 \times 10^9$ $^{87}Rb$ atoms or $7 \times 10^7$ $^{40}K$ atoms. Loading rates as high as $1.2 \times 10^9$ $^{87}Rb$ atoms/s and $8 \times 10^7$ $^{40}K$ atoms/s were achieved without the use of a secondary atom source. After the desorption light is turned off, the pressure quickly decays back to equilibrium with a  time constant as short as $200\ \mu s$, allowing for long trapping lifetimes after the MOT loading phase.

\end{abstract}

\maketitle
\par

\section{Introduction }

In recent years the field of ultracold atomic gases has made a number of significant contributions to our understanding of atomic and molecular physics as well as precision metrology. Spectacular experiments included the observation of Bose-Einstein condensation (BEC)~\cite{And95,Dav95}, quantum degenerate Fermi gases~\cite{DeM99_1, Tru01, Sch01, Gra02, Roa02, OHa02, Had03} and more recently the BEC-BCS transition in fermionic gases~\cite{Reg04, Zwi04, Kin04, Bar04, Kin05}. New sensors and clocks based on new cooling methods are on the horizon \cite{Tak05,Stu03,Arl05}. The development of a number of important tools such as optical dipole traps~\cite{Gri00}, optical lattices~\cite{Blo05} and magnetic control of interatomic potentials~\cite{Ino98} has fostered these rapid advances in the field. All of the fascinating experimental results rely on the development of laser cooling and trapping~\cite{met99} for neutral atoms in the 1980s. This work led to the realization of the first magneto-optical trap (MOT)~\cite{Raa87} which is now routinely used to trap up to a few $10^9$ atoms at $\mu K$ temperatures. So far almost all experiments with ultracold atoms rely on the MOT as a first step to collect and cool atomic clouds.

However, there is a basic limitation that all experiments have to overcome. For the loading of a trap, a large particle flux is required while long lifetimes of the sample are desired for efficient evaporative cooling towards quantum degeneracy as well as long drift times for atom interferometry. Since the gain mechanism often also induces a particle loss, these two requirements limit each other.

A number of solutions have been used to overcome this problem. Initial BEC experiments showed that there is a compromise that allows for the production of small BECs. Larger BECs were produced by using an atomic beam that can be shut off after loading of the MOT. The next generation of systems for the production of BEC relied on two MOTs, where one MOT is used to trap atoms in a region of high background pressure and the precooled atoms are then transferred to a region of better vacuum where they are captured by a second MOT. Many of the most recent systems employ a purely magnetic transfer mechanism to move precooled atoms to a region of better vacuum \cite{Gre01,Lew03}. Thus, only a single MOT is required, allowing for improved optical access for further experiments.

Two recently developed techniques have shown that there is another possible path towards BEC. The large confinement that can be reached by atom chips and optical traps allows for extremely rapid evaporative cooling and therefore places less stringent requirements on the background pressure.

Despite these advantages, most of these experiments will profit from low background pressures for the actual experiments after the MOT phase. Two techniques can help to change the partial pressure of the desired atomic species on short timescales. First, a so called dispenser can be used to provide the desired atomic species by reducing an alkali salt. Such dispensers are commercially available for some elements and can be assembled for some others~\cite{DeM99}. By ohmic heating of such a dispenser the partial pressure of the atomic gas can be raised during the MOT loading phase and subsequently drops when the current is shut off to allow for a better vacuum for further experiments \cite{For98}. In this situation, the pumping speed of the vacuum system determines the decay time of the pressure. Unfortunately, this time constant is usually long in comparison to the experiment cycles.

A second technique available for this purpose is light-induced atom desorption. Atoms that are adsorbed at the walls of a vacuum chamber can be desorbed by irradiation with weak and incoherent light. This  allows for a temporary increase of the desired partial pressure. LIAD has been investigated for a number of alkali atoms since 1993~\cite{goz93}. However, most investigations have focused on spin polarized room temperature gases in coated quartz cells. For the case of rubidium atoms it has also been used to load a rubidium MOT~\cite{and01} and it has been shown to allow for high loading efficiencies in a MOT~\cite{atu03}.

We have extended this work by evaluating the wavelength and intensity dependence of LIAD for a dual species MOT for $^{87}Rb$ and $^{40}K$ in an uncoated quartz cell. With UV desorption light intensities smaller than $25\ mW/cm^2$, the obtained partial pressures were indeed high enough to saturate the atom number in the MOTs without the use of  a secondary atom source. In our case, the MOTs saturate at $3.5 \times 10^9 $ rubidium atoms and $7 \times 10^7$ potassium atoms. We show that the pressure quickly returns to equilibrium after the desorption light is turned off. This allows for long storage times in a quadrupolar magnetic trap and a low-loss transfer of the atoms into another region for further experiments.

In a second experimental setup based on a stainless steel chamber, we also show that LIAD with UV light can serve as an atomic source for magneto-optical traps. In this case, the desorption light intensity can be reduced by four orders of magnitude in comparison with white desorption light~\cite{and01}.

\section{Background}
\label{background}

This section briefly reviews the basic properties of the magneto-optical trap and light-induced atom desorption. These properties will be used for a further analysis of the results in section~\ref{results}.

\subsection*{MOT dynamics}

A vapor-cell MOT is designed to capture atoms from the background gas. Atoms below a critical velocity are slowed by laser light and collected at the center of a magnetic quadrupole field. The number of collected atoms $N(t)$ is governed by a rate equation~\cite{mon90}

\be
\frac{dN}{dt} = R -
N\left(\frac{1}{\tau_{1}}+\frac{1}{\tau_{2}}\right)- \beta \int n^2
dV. 
\label{eq:rateequation} 
\ee

The atom number gain is characterized by a loading rate $R$. This rate is defined by the technical properties of the MOT (laser beams and magnetic field). It is directly proportional to the partial pressure of the considered atomic species at a given temperature. Higher temperatures reduce the fraction of slow atoms and consequently the amount of trappable atoms. Collisions between trapped atoms and atoms in the background gas induce a loss of atoms which depends on the number of trapped atoms. $N/\tau_1$ represents the loss rate due to collisions with untrapped atoms of the same species whereas $N/\tau_2$ denotes the loss due to collisions with all other species in the background gas. As the number of trapped atoms rises, density dependent collisions become more dominant. The number of collisions in the MOT is proportional to the  integral on the right. The proportionality is expressed by the loss coefficient $\beta$. If the density dependent loss is
neglected, the rate equation can be easily solved by

\be N(t)=N_{max}\left(1-e^{-\frac{t}{\tau_{MOT}}}\right),
\label{eq:loading} \ee

where the MOT loading time is defined by $1/\tau_{MOT}=1/\tau_1+1/\tau_2$ and $N_{max}$ represents the maximum atom number. The loading rate

\be R=\frac{dN}{dt}(0)=\frac{N_{max}}{\tau_{MOT}}  \label{eq:rate} \ee

can thus be used to measure the partial pressure of the considered atom.

\subsection*{LIAD}

The effect of light-induced atom desorption has been observed for $Rb$, $Cs$, $Na$, $Na_2$, $Ni$, $Zn$, $Sn$ and $K$ atoms, using various cell materials (Pyrex, porous silica, sapphire and stainless steel), mostly coated with polydimethylsiloxane (PDMS), octamethylcyclotetrasiloxane (OCT) or paraffin~\cite{goz93, abr84, meu94, mar94, xu96, atu99, ale02, goz04, and01, bur04}.

In particular the  wavelength dependence has been investigated with $Na$, $Na_2$~\cite{xu96}, $K$ \cite{goz04} and $Rb$~\cite{meu94} for PDMS coating and with $Rb$ and $Cs$~\cite{ale02} for paraffin coating. However, a measurement of the wavelength dependence with an uncoated cell has been outstanding. Desorption light of different wavelengths was compared at the same intensity and it was shown that desorption starts above a characteristic energy threshold and the atom yield grows quadratically with the photons' energy. This relationship can be described in analogy to the well-understood photoelectric effect~\cite{xu96}.

Higher intensities of desorbing light lead to a higher flux as more atoms are desorbed from the walls of the cell. The flux grows linearly at low intensities, but increases only with a square root dependence for higher intensities~\cite{atu99,goz04,bur04}. A saturation of the desorption yield for high intensities was reported in Ref.~\cite{meu94}. Reference \cite{atu99} proposes rate equations which reproduce the linear and the square root dependence. The diffusion of Rb atoms inside the PDMS coating is considered to be the dominant process. However, this model is only applicable to systems with coated cells and a description for uncoated cells is outstanding.

\begin{figure}[t]

\centering

\includegraphics*[width=8.6cm]{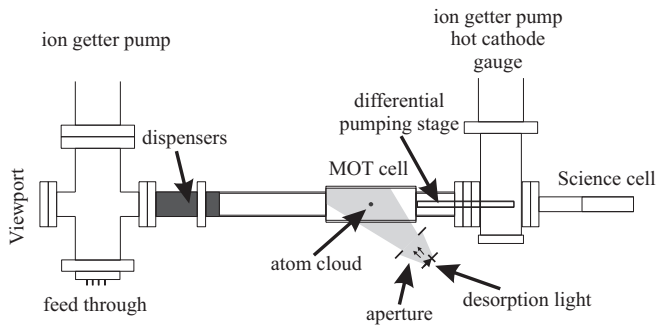}

\caption{Vacuum system of the experimental apparatus. The position of the dispensers and the light LIAD light are indicated.}

\label{fig:vacuum}

\end{figure}

\section{Experimental Setup}
\label{setup}

The LIAD studies were performed in two different experimental setups. Most of the results were obtained in an apparatus designed for the production of quantum degenerate K-Rb mixtures which is described in this section. The second apparatus based on a stainless steel chamber is designed to produce a $^{87}Rb$ Bose-Einstein condensate on a chip (described in detail in Ref.~\cite{Vog05}). The design of these experiments motivated the reported investigation of LIAD.

\subsection*{MOT design}

Figure~\ref{fig:vacuum} shows the vacuum system which constitutes the core of our experimental apparatus. It consists of two chambers separated by a differential pumping stage. The larger cell contains a MOT for collection of atomic samples (MOT cell) and the smaller one allows for evaporative cooling in a magnetic trap (science cell).

Both cells consist of UV transmitting quartz~\cite{vycor}. The inner surface of the MOT cell has the dimensions $50\ mm \times 50\ mm \times 140\ mm$ ($5\ mm$ thickness). Both cells are pumped by ion getter pumps and the good vacuum in the science cell is sustained by a supplementary titanium sublimation pump. Typically pressures of $1 \times 10^{-9} \ mbar$ and $2 \times 10^{-11} \ mbar$ respectively are reached in the two vacuum regions. We use commercial rubidium dispensers and potassium dispensers constructed according to Ref.~\cite{DeM99} to provide vapors of $^{87}Rb$ and $^{40}K$. These dispensers are located at a distance of $\approx 35\ cm$ from the MOT cell and are used to coat the surfaces with a rubidium-potassium-mixture.

Two laser systems are necessary to provide the light for magneto-optical trapping of the two atomic species. In both cases high laser powers are required for trapping large atomic samples. The cooling and repumping light for the rubidium atoms is provided by two external cavity diode lasers. The cooling light is tuned 24 MHz below the $5S_{1/2}, F=2 \rightarrow 5P_{3/2}, F'=3$ transition and the repumping light is resonant with the $5S_{1/2}, F=1 \rightarrow 5P_{3/2}, F'=2$ transition. Both beams are superposed and simultaneously amplified by a tapered amplifier chip~\cite{TA}. A Titanium-Sapphire laser provides both cooling and repumping light for the potassium atoms. The cooling light is shifted 24 MHz below the $4S_{1/2}, F=\frac{9}{2} \rightarrow 4P_{3/2}, F'=\frac{11}{2}$ transition by an acousto-optical modulator (AOM) in double-pass configuration. The repumping light is tuned resonant to the $4S_{1/2}, F=\frac{7}{2} \rightarrow 4P_{3/2}, F'=\frac{9}{2}$ transition by an AOM in quadruple-pass configuration. In total, this setup provides $190 \ mW$ ($118 \ mW$) of cooling power and $10 \ mW$ ($26 \ mW$) of repumping power for rubidium (potassium) atoms. All beams are combined using polarizing beam splitters and long pass mirrors~\cite{longpass} before being coupled into a single polarization maintaining fiber. The use of a single fiber for transporting all trapping light to the experiment greatly facilitates all further adjustments.

After the fiber, the optics is not more complicated than for a single species MOT. The fiber output is split into 6 beams, expanded to $3 \ cm$ in diameter, circularly polarized and directed onto the atomic cloud by dichroic optics. The maximum intensity is $6.5 \ mW/cm^2$ ($4.8 \ mW/cm^2$) per beam for rubidium (potassium) light. The magnetic field gradient of $9.2 \ G/cm$ is provided by two coils in anti-Helmholtz configuration. These coils can produce a quadrupole field with a gradient of up to $138\ G/cm$ which can be used to trap the atomic samples magnetically for further experiments \cite{Gre01,Lew03}. This configuration is designed for particularly large vapor-cell MOTs, since the number of collected atoms cannot be increased after the transport to the science cell. 

For comparative measurements, a mirror MOT in a steel chamber was used~\cite{Vog05}. The apparatus for the production of a $^{87}Rb$-BEC on a chip has an extremely compact and robust design for operation during free fall from a droptower. Due to the shape of the chamber, the optical access for LIAD is greatly reduced, but allows for a comparison between LIAD from a quartz and a steel surface for $^{87}Rb$ atoms.

\subsection*{LIAD benefits}
\label{translation}

The work on using LIAD as an atomic source was motivated on one hand by our magnetic transport mechanism, since losses due to collisions with the background gas can be reduced by using the LIAD effect. On the other hand, experiments in free fall impose strict requirements on the complexity of the vacuum system.

After the desorption light is turned off, the pressure decays back to equilibrium. The decay time depends only on the specific properties of the adsorption process. Reference~\cite{and01} empirically discovered a double exponential decay. A possible explanation for the two time constants $\tau_{short}$ and $\tau_{long}$ is due to the adsorption properties of the cell walls. As long as the cell's surface is not completely covered by the atomic species, the time constant $\tau_{short}$ is produced by quartz-atom adsorption. Once the cell is completely covered, the much weaker atom-atom adsorption governs the pressure evolution by $\tau_{long}$.

This behavior implies a three-phase strategy for experiments using a LIAD source~\cite{and01}: The MOT is loaded with desorption light on, it is then held without desorption light while the pressure drops and finally, only magnetic or optical fields are used to confine the atoms.

\subsection*{Desorption experiments}

To investigate the influence of desorption light, the appropriate dispenser was fired for a few hours to coat the quartz cell with rubidium or potassium. Afterwards, the system was pumped for a few days to restore the former  equilibrium pressure of $~ 1\times 10^{-9} \ mbar$. We then obtained only negligible MOT sizes ($< 5\times10^7$ $^{87}Rb$ atoms and $ < 1\times10^6$ $^{40}K$ atoms) without the use of LIAD.

Desorption light at 5 different wavelengths was used in our experiments. Red ($630 \ nm$), green ($520 \ nm$), blue ($470 \ nm$) and UV-A ($395 \ nm$) light was generated by powerful LED arrays~\cite{LEDmanufacturer}. The UV-C light at $253~nm$ was generated by a fluorescent tube from an EPROM eraser~\cite{eprom}. To compare the LIAD effect at different wavelengths with similar irradiation, we mounted all lamps in the same position, at a distance of $8 \ cm$ from a $5 \ cm$ aperture as shown in Fig.~\ref{fig:vacuum}. The LEDs were operated at an intensity of $1 \ mW/cm^2$, measured at a distance of $17,5 \ cm$, comparable to the distance between the light sources and the cell. Since the UV-C tube only produces a maximum intensity of $150 \ \mu W/cm^2$, the UV-A LED was also operated at $120 \ \mu W/cm^2$ to allow for a comparison of the two. The difference of $30 \ \mu W/cm^2$ compensates the smaller UV-C transmission coefficient of the used Vycor glass.

The intensity dependence was investigated by placing four arrays (consisting of eight $395 \ nm$-LEDs each) close to the MOT cell. This results in both higher light intensities and a larger irradiated cell surface.

\subsection*{Atom detection}

All results were obtained by monitoring the atom number during the loading of the MOT. Due to the substantially different atom numbers obtained for $^{87}Rb$ and $^{40}K$ atoms, two different fluorescence imaging techniques are used.

The rubidium cloud is imaged onto a large area photodiode by a single lens. The resulting signal is directly proportional to the atom number for a given set of experimental parameters. The potassium cloud is monitored by a CCD camera. Both the pixel sum of the CCD camera and the photodiode current were calibrated and checked for consistency. Thus, the respective atom numbers can be inferred with an estimated error of $30\%$. However, most results presented here depend on the loading rate rather than the absolute atom number. Therefore, much higher precision is obtained in those measurements.

\begin{figure}[!ht]

\centering

\includegraphics*[width=8.6cm]{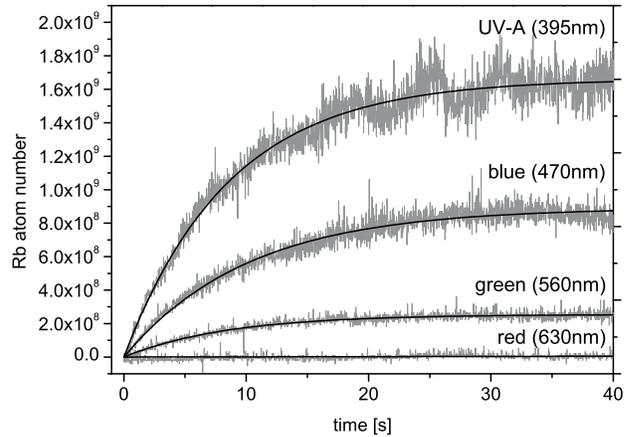}

\caption{$^{87}Rb$ atom number as a function of the MOT loading time for desorption light at various wavelengths.}

\label{fig:loading}

\end{figure}

\section{Results}
\label{results}

This section investigates the dependence of the LIAD effect on the desorption wavelength and its intensity. By monitoring the pressure in the system, we demonstrate the benefits of LIAD for experiments towards quantum degenerate atomic gases. 

\subsection*{LIAD wavelength dependence}

The time evolution of the atom number during the loading phase of the MOT is shown in Fig.~\ref{fig:loading} for typical realizations. In this case, the $^{87}Rb$ atom number is displayed for various wavelengths of the desorption light. For desorption light at $630\ nm$, the atom number is below the noise limit, equivalent to the case without any desorption light. The maximum atom numbers and the loading time constants were extracted from fits according to Eq. \ref{eq:loading}. These fits match the data very well for all investigated configurations and the relative statistical errors of the fit parameters are well below $1\%$. Therefore, no error bars are displayed in the following figures.

\begin{figure}[!ht]

\centering

\includegraphics*[width=8.6cm]{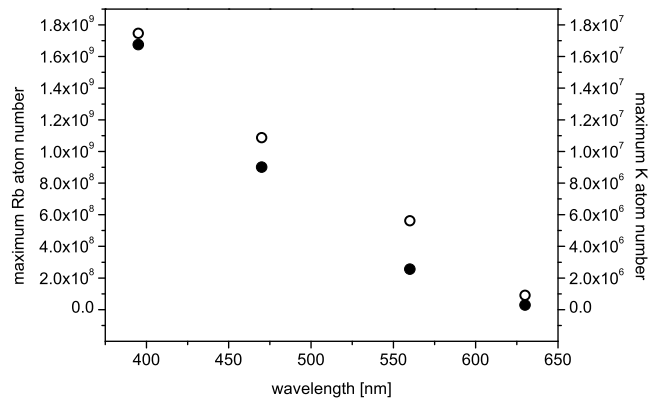}

\caption{Maximum atom number in the MOT versus desorption light wavelength, for $^{87}Rb$ (solid dots) and $^{40}K$ (open circles).}

\label{fig:atomnumber}

\end{figure}

Figure \ref{fig:atomnumber} shows the maximum atom numbers in the rubidium and the potassium MOT depending on the wavelength of the incident desorption light. Both for $^{87}Rb$ and $^{40}K$, a substantial increase in the atom number is observed towards shorter desorption wavelengths.

\begin{figure}[!ht]

\centering

\includegraphics*[width=8.6cm]{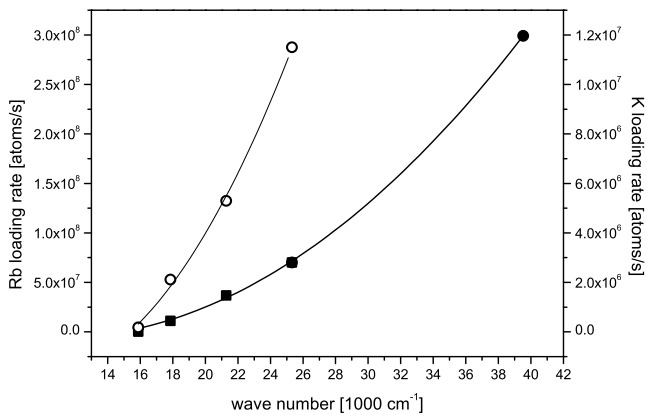}

\caption{$^{87}Rb$ (solid symbols) and $^{40}K$ (open circles) loading rates versus wave number of the desorption light. The solid squares for $^{87}Rb$ represent rescaled data (see text).}

\label{fig:rate}

\end{figure}

However, the loading rate is more suitable to quantify the desorption efficiency, since it is proportional to the partial pressure of the atomic species. In this case, it is more instructive to plot the loading rate as a function of the wave number instead of the wavelength. Figure \ref{fig:rate} shows the loading rates according to Eq. \ref{eq:rate} for both potassium and rubidium atoms. The loading rates were measured for desorption with red, green, blue and UV-A light for $^{87}Rb$ and $^{40}K$. This figure is complimented by an additional measurement for $^{87}Rb$ with desorption light at $253\ nm$. Since this data point was recorded at lower intensity, the data at longer wavelengths were rescaled accordingly. Similarly to the atom number, the loading rates are increased substantially for lower desorption wavelengths.

We observe a threshold in photon energy of the desorption light, similarly to the case of LIAD from sapphire or coated surfaces. The desorption for $630 \ nm$ and the laser light at $780 \ nm$ ($767 \ nm$) is below our noise limit. The observed threshold is shifted to higher values in comparison to other materials which is due to the specific adsorption energy.

A parabolic fit was used to infer the exact energy thresholds. We found  a $^{87}Rb$ threshold at a wave number of $14800 \pm 1900\ cm^{-1}$ (corresponding to $675 \pm 85 \ nm$) and a $^{40}K$ threshold at a wave number of $15400 \pm 700 \ cm^{-1}$ (corresponding to $651 \pm 29 \ nm$). Both thresholds are higher than the ones measured with PDMS coatings~\cite{goz04}. However, a microscopic analysis of these thresholds is beyond the scope of this paper.

In summary, our measurements show that LIAD with short wavelength light is ideally suited as an atomic source of $^{87}Rb$ and $^{40}K$.

\subsection*{LIAD intensity dependence}

\begin{figure}[!ht]

\centering

\includegraphics*[width=8.6cm]{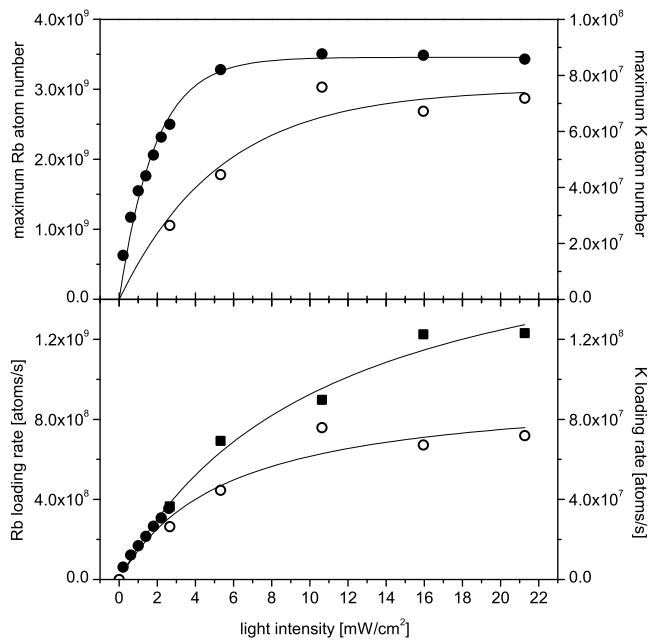}

\caption{$^{87}Rb$ (solid symbols) and $^{40}K$ (open circles) atom numbers (top) and loading rates (bottom) versus desorption light intensity. The model which was used to fit the data will be evaluated in future work.}

\label{fig:intensity}

\end{figure}

The intensity dependence of the LIAD effect was studied at a desorption wavelength of $395\ nm$. This wavelength was chosen for further experiments since it is sufficiently short to produce a large LIAD effect without resulting in health risks. The MOT loading rate and total atom number were measured as a function of the light intensity supplied by four LED arrays. Figure \ref{fig:intensity} shows the MOT loading rate for $^{87}Rb$ and $^{40}K$ as a function of the desorption intensity. The $^{87}Rb$ data points at low intensities (solid dots) were recorded with only one LED array. To calibrate the intensity, a smooth transition at $2.6 \ mW/cm^2$ was assumed.

The loading rates for both $^{87}Rb$ and $^{40}K$ rise significantly with increasing desorption light intensity. We obtained maximum loading rates of $1.2 \times 10^9$ $^{87}Rb$ atoms/s and $8 \times 10^7$ $^{40}K$ atoms/s. The employed intensities are sufficient to saturate the atom numbers in both MOTs at $3.5 \times 10^9 $ rubidium atoms and $7 \times 10^7$ potassium atoms (see top of Fig. \ref{fig:intensity}) without the use of a secondary atom source. This means that the dispensers do not have to be used regularly and that the experiment can rely entirely on the use of LIAD. Therefore, most of the experiments can be conducted at the optimal pressure obtained in the vacuum system.

\subsection*{MOT loss rate}

To evaluate the use of LIAD as an atom source, it is necessary to check if the desorption of unwanted atomic species results in enhanced loss rates. One important contamination originates from the untrapped $Rb$ and $K$ isotopes liberated by the dispensers ($^{85}Rb$: $72.17 \%$ and $^{39}K$: $63.45 \%$, $^{41}K$: $29.05 \%$). LIAD will desorb all these isotopes from the cell walls. Therefore, the $^{87}Rb$ partial pressure cannot be varied independently by LIAD and a clear distiction between the loss processes is not possible. However, the loss rates for the MOT operation with dispensers can be compared to the loss rates with LIAD. This allows for an evaluation of the relative contamination of the vacuum by the two atom sources.

The inverse loading time $1/\tau_{MOT}$ represents the MOT loss rate due to collisions with the background gas (see Eq. \ref{eq:rateequation}). Although its dependence on the partial pressures is complex, it monotonically increases with the contamination of the vacuum system. In order to probe this contamination, the inverse loading time and the loading rate were extracted from fits according to Eqs. \ref{eq:loading} and \ref{eq:rate}. The inverse loading time was compared at the same loading rate (i.e. the same $^{87}Rb$ partial pressure) for dispenser and LIAD operation.

The MOT operation results in approximately equal loss rates with both atom sources. The inverse loading time with LIAD is $6.5\% \pm 5.1\%$ lower than for dispenser operation. This means that the background gas produced by LIAD is slightly less contaminated than the dispenser output. Therefore, the MOT saturates at a slightly higher maximum atom number with LIAD at equal $^{87}Rb$ partial pressure. A possible explanation for this improvement is that some contaminants produced by the dispensers are adsorbed tightly to the cell walls. Since LIAD does not produce enhanced MOT loss rates, it serves as a better atom source in comparison to dispensers.

\subsection*{Pressure decay}

To be able to benefit from the pressure modulation by LIAD for subsequent experiments, it is necessary that the pressure decays quickly after the desorption light is turned off.

\begin{figure}[!ht]

\centering

\includegraphics*[width=8.6cm]{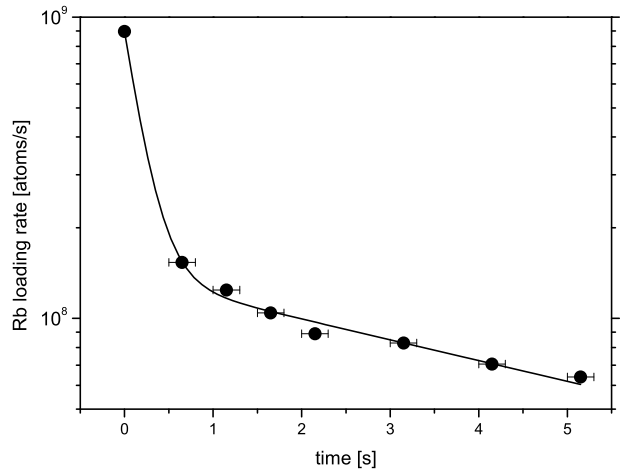}

\caption{$^{87}Rb$ loading rate as a function of the delay after the desorption light was turned off.}

\label{fig:pressure}

\end{figure}

To measure the relevant timescales, the MOT loading was evaluated at different times after the desorption light was turned off. Figure \ref{fig:pressure} shows the loading rate as a function of this delay. The loading rates were measured by evaluating the atom number gain in the MOT within a $300\ ms$ interval. Note that the first data point corresponds to the loading rate at $10.6\ mW/cm^2$ from Fig. \ref{fig:intensity}.

As described in section \ref{setup}, we fitted this data with a double exponential decay in order to extract the two relevant time scales. The fast component $\tau_{short} = 200 \pm 20 \ ms$ and the slow component $\tau_{long} = 6.3 \pm 0.9 \ s$ were thus extracted. As expected, the initial decay of the pressure after the use of LIAD is extremely fast and restores good vacuum conditions rapidly. In fact, the pressure drops by a factor of 7.5 in the first second. It is therefore beneficial to hold the MOT during this fast decay. We have measured a loss of only $ 6.4\%$ of the atoms during this time. At this point, further experiments in a magnetic trap drastically benefit from a lifetime which is increased by roughly an order of magnitude. Thus, all experiments using a vapor-cell MOT can be easily improved by using LIAD as a switchable atom source.

\begin{figure}[!htb]

\centering

\includegraphics*[width=8.6cm]{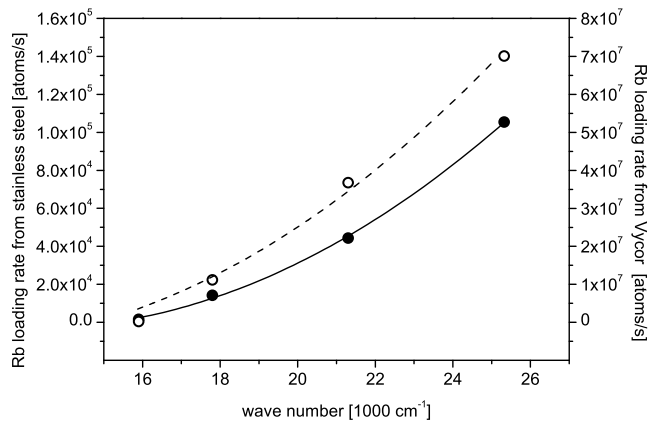}

\caption{$^{87}Rb$ loading rate versus wave number of the desorption
light in a stainless steel chamber (solid dots) and a quartz cell (open circles).}

\label{fig:stainless}

\end{figure}

\subsection*{LIAD in a stainless steel chamber}

In order to demonstrate the benefits of LIAD for further types of experiments, we complimented our investigation with a measurement of the atom desorption in an uncoated stainless steel chamber. Since the magneto-optical trap in this chamber only has $2\ cm$-beams with a maximum intensity of $4\ mW/cm^2$, the resulting MOT loading rates are much smaller compared to the measurements presented so far. The LIAD measurements were performed with desorption light at the same choice of wavelengths and a constant intensity ($\approx 1\ mW/cm^2$) analogous to the first subsection. Figure \ref{fig:stainless} shows the MOT loading rate as a function of the desorption wave number in both the steel chamber and the quartz cell. The desorption yield in the steel chamber is clearly enhanced by applying light at short wavelengths. The extracted threshold at $15200 \pm 800 \ cm^{-1}$ (corresponding to $656 \pm 35 \ nm$) is remarkably similar to result for quartz. It is striking that the maximum loading rate was only a factor of 8 smaller than the one reported for white light~\cite{and01}, although the desorption light intensity was decreased by four orders of magnitude ($\approx 1\ mW/cm^2$ versus $\approx 10\ W/cm^2$).

\section{Summary and Outlook}

In conclusion, we have measured the wavelength and intensity dependence of LIAD for both $^{87}Rb$ and $^{40}K$ from a quartz and a stainless steel surface. We have shown that desorption light at low intensities can be used to produce sufficient vapor pressures to operate large magneto-optical traps at $3.5 \times 10^9$ $^{87}Rb$ atoms or $7 \times 10^7$ $^{40}K$ atoms. Loading rates as high as $1.2 \times 10^9$ $^{87}Rb$ atoms/s and $8 \times 10^7$ $^{40}K$ atoms/s were achieved without the use of a secondary atom source. We measured the wavelength dependence of the LIAD effect and extracted the relevant energy thresholds. Using desorption as a switchable atom source, we demonstrated a technique to obtain significantly reduced atom loss in trapping experiments. In particular, this technique can be used to simplify BEC setups and to realize atom chip experiments with long observation times.

A number of aspects remain to be addressed in future work. It is necessary to develop a microscopic model of the LIAD process which explains the measured dependencies as a surface effect, since the explanation of Ref.~\cite{atu99} is only applicable to volume diffusion inside a coating. Further experiments may obtain much higher atom numbers in a vapor-cell MOT using the following approach: It has been shown that LIAD works in glass cells which are held at liquid helium temperatures~\cite{hat02}. In these conditions, a much larger fraction of the atoms in the background gas can be captured by a MOT, resulting in extremely enhanced loading rates. Since the maximum atom number in a MOT is limited by the ratio of loading to loss rate, magneto-optical traps with much higher atom numbers will be achievable. \\

\section{Acknowledgments}

We thank Klaus Sengstock, Kai Bongs, Silke and Christian Ospelkaus from the ``Institut f\"ur Laser-Physik'' in Hamburg and Prof. Binnewies from the ``Institut f\"ur Anorganische Chemie'' in Hannover for their help with the preparation of the potassium dispensers. We thank M. Volk for stimulating discussions. We acknowledge support from the Deutsche Forschungsgemeinschaft (SFB 407), the European Graduate College Orsay-Glasgow-Hannover and the Deutsche Luft- und Raumfahrtgesellschaft (DLR 50 WM 0346).


\begin{thebibliography}{99}
\bibitem{longpass} The employed long pass mirrors reflect $> 95\%$ of $767\ nm$ laser beams and transmit $> 95\%$ of $780\ nm$ beams. The long pass mirrors were manufactured by the Laser Zentrum Hannover e.V..
\end{thebibliography}
\end{document}